\documentclass[aps,prc,twocolumn,showpacs,floatfix,nofootinbib]{revtex4}
\usepackage{amsmath,graphicx}

\begin{document}
\preprint{}
\title{Particle production in relativistic heavy-ion collisions: 
A consistent hydrodynamic approach}
\author{Rajeev S. Bhalerao, 
Amaresh Jaiswal,
Subrata Pal and
V. Sreekanth}
\affiliation {Tata Institute of Fundamental Research,
Homi Bhabha Road, Mumbai 400005, India}

\date{\today}

\begin{abstract}
We derive relativistic viscous hydrodynamic equations invoking the
generalized second law of thermodynamics for two different forms of
the non-equilibrium single-particle distribution function. We find
that the relaxation times in these two derivations are identical for
shear viscosity but different for bulk viscosity. These equations are
used to study thermal dilepton and hadron spectra within longitudinal
scaling expansion of the matter formed in relativistic heavy-ion
collisions. For consistency, the same non-equilibrium distribution
function is used in the particle production prescription as in the
derivation of the viscous evolution equations. Appreciable differences
are found in the particle production rates corresponding to the two
non-equilibrium distribution functions. We emphasize that an
inconsistent treatment of the non-equilibrium effects influences the
particle production significantly, which may affect the extraction of
transport properties of quark-gluon plasma.
\end{abstract}



\maketitle

\section{Introduction}
Evolution of the strongly-interacting matter produced in high-energy
heavy-ion collisions, when the system is close to local thermodynamic
equilibrium, has been studied extensively within the framework of the
relativistic dissipative hydrodynamics; for a recent review see
Ref. \cite{Heinz:2013th}. As the system expands and becomes dilute
enough, the hydrodynamic description breaks down, leading to a
freezeout or a transition from the hydrodynamic description to a
particle description \cite{Cooper:1974mv}. The dissipative effects are
important not only during the hydrodynamic evolution, but also in the
particle production \cite{Teaney:2003kp}, and the two have to be treated
in a consistent manner. Moreover, the transport coefficients and
relaxation times which constitute an external input to the
hydrodynamic equations need to be in conformity with the theoretical
framework used to derive the hydrodynamic equations
\cite{Jaiswal:2013fc}. {\it Ad hoc} choices or inconsistent treatments could
significantly affect the final-state particle yields, spectra, and
other observables derived from them.

Hydrodynamics is formulated as an order-by-order expansion in
gradients of the hydrodynamic four-velocity $u^\mu$. The ideal
hydrodynamics is zeroth order and relativistic Navier-Stokes theory is
first order in gradients; the latter violates causality. Derivation of
the (causal) second-order dissipative hydrodynamic equations proceeds
in a variety of ways; for a review see Ref. \cite{Romatschke:2009im}. For
instance, in the derivations based on kinetic theory the
non-equilibrium phase-space distribution function $f(x,p)$ needs to
be specified. This is commonly achieved by taking recourse to Grad's
14-moment approximation \cite{Grad}. The hydrodynamic equations are
then derived by suitable coarse-graining of the microscopic
dynamics. For consistency, the same $f(x,p)$ ought to be used in the
particle-production prescription \cite{Cooper:1974mv,McLerran:1984ay}
as well. This important consideration has been overlooked in several
hydrodynamic studies of heavy-ion collisions.

An alternate derivation of hydrodynamic equations starts from a
generalized entropy four-current $S^\mu$, expressed in terms of a few
unknown coefficients, and then invokes the second law of thermodynamics
($\partial_\mu S^\mu \geq 0$) \cite{Romatschke:2009im}. These
coefficients which are related to relaxation times for shear and bulk
pressures remain undetermined, and have to be obtained from kinetic
theory \cite{Israel:1979wp,Muronga:2003ta}. Even then the bulk
relaxation time remains ambiguous. Ideally, a single theoretical
framework should give rise to dissipative evolution equations as well
as determine these unknown coefficients \cite{Jaiswal:2013fc}. The
bulk relaxation time obtained in Ref. \cite{Jaiswal:2013fc} exhibits
critical slowing down near the QCD phase transition and does not lead
to cavitation.

In this paper, we employ the method based on the entropy four-current
to derive second-order viscous hydrodynamics corresponding to two
different forms of the non-equilibrium distribution function. These
distribution functions are formally different, and one of them is used
here for the first time to study the particle production in heavy-ion
collisions. We perform a comparative numerical study of these two
formalisms in the Bjorken scaling expansion. For consistency, we use
the same non-equilibrium distribution function in the calculation of
the particle spectra as in the derivation of the evolution equations,
and point out drawbacks of an inconsistent treatment. As an
application, we study the production of thermal dileptons and hadrons
in various scenarios.

The paper is organized as follows. In Sec. II, we derive viscous
hydrodynamic equations corresponding to two choices of the
non-equilibrium distribution function. In Sec. III we present thermal
dilepton and hadron production rates. In Sec. IV we consider the
Bjorken scenario and obtain the relevant evolution equations and
particle production rates in the two cases. Numerical results are
presented and discussed in Sec. V, which is followed by the Summary in
Sec. VI.


\section{Viscous hydrodynamics}

The entropy four-current for particles obeying the Boltzmann
statistics is given by \cite{deGroot}
\begin{equation}\label{EFC}
S^\mu(x) = -\int dp ~p^\mu f \left(\ln f - 1\right),
\end{equation}
where $dp = g d{\bf p}/[(2 \pi)^3\sqrt{{\bf p}^2+m^2}]$, $g$ and $m$
being the degeneracy factor and the particle rest mass, respectively,
$p^{\mu}$ is
the particle four-momentum, and $f\equiv f(x,p)$ is the
single-particle phase-space distribution function. For a system close
to equilibrium, $f$ can be written as $f = f_0 + \delta f \equiv
f_0(1+\phi)$, where the equilibrium distribution function is defined
as $f_0 = \exp(-\beta u\cdot p)$. Here $\beta \equiv 1/T$ is the
inverse temperature, $u^\mu$ is defined in the Landau frame
\cite{deGroot}, and we have assumed the baryo-chemical potential to be
zero.

The divergence of $S^\mu$ reads
\begin{align}\label{EFCD}
\partial_\mu S^\mu &= -\int dp ~p^\mu \left(\partial_\mu f\right) 
\ln f \nonumber \\
&= -\int dp~p^\mu \left[\phi(1+\phi/2)(\partial_\mu f_0)+\phi(\partial_\mu \phi)f_0\right],
\end{align}
where in the second equality terms up to third order in gradients have
been retained.

To proceed further, the non-equilibrium part of the distribution
function $\delta f \equiv f_0\phi$ needs to be specified. In the
present study, we consider two different forms of $\phi$. The first
form is consistent with Grad's 14-moment approximation \cite{Grad} for
the single-particle distribution function in orthogonal basis \cite
{Denicol:2012cn}. We propose \cite{Jaiswal:2013fc}
\begin{equation}\label{phi1}
\phi_1 = \frac{\Pi}{P} + \frac{p^\mu p^\nu \pi_{\mu\nu}}{2(\epsilon+P)T^2}, 
\end{equation}
where corrections up to second order in momenta are present. Equation
(\ref{phi1}) has not been used before to study particle production in
heavy-ion collisions. The second form is obtained by considering the
corrections which are only quadratic in momenta \cite{Dusling:2007gi}:
\begin{equation}\label{phi2}
\phi_2 = \frac{p^\mu p^\nu}{2(\epsilon+P)T^2}\left(\pi_{\mu\nu}
+\frac{2}{5}\Pi\Delta_{\mu\nu}\right).
\end{equation}
In Eqs. (\ref{phi1}) and (\ref{phi2}), $\epsilon$ and $P$ are the
thermodynamic energy density and pressure, respectively,
$\Pi$ the bulk viscous
pressure, $\pi^{\mu\nu}$ the shear stress tensor, and
$\Delta^{\mu\nu}=g^{\mu\nu}-u^\mu u^\nu$. The energy-momentum tensor
can be expressed in terms of these quantities as $T^{\mu\nu} =
\epsilon u^\mu u^\nu-(P+\Pi)\Delta^{\mu \nu} + \pi^{\mu\nu}$. Note
that although the contributions due to shear in Eqs. (\ref{phi1}) and
(\ref{phi2}) are identical, those due to bulk viscosity are distinct.
In the following, we shall refer to analyses performed using
Eqs. (\ref{phi1}) and (\ref{phi2}) as `Case 1' and `Case 2',
respectively.

Performing the integrals in Eq. (\ref{EFCD}) as outlined in
Ref. \cite{Jaiswal:2013fc}, we obtain
\begin{align}\label{EFCD3}
\partial_\mu S^\mu = 
& - \beta\Pi\left[ \theta +\! \beta_0\dot\Pi 
+ \frac{4}{3}\beta_0 \theta\Pi \right] \nonumber \\
&+ \beta\pi^{\mu\nu}\left[ \sigma_{\mu\nu} 
- \beta_2\dot\pi_{\mu\nu} 
- \frac{4}{3}\beta_2 \theta\pi_{\mu\nu} \right],
\end{align}
where $\beta_0$ and $\beta_2$ are functions of thermodynamic
quantities $\epsilon$ and $T$, $\dot X \equiv u^\mu \partial_\mu X$,
$\theta=\partial_\mu u^\mu$, and $\sigma^{\mu\nu}=\nabla^{\langle \mu}
u^{\nu \rangle}$. The notation $A^{\langle\mu\nu\rangle} =
\Delta^{\mu\nu}_{\alpha\beta}A^{\alpha\beta}$, where
$\Delta^{\mu\nu}_{\alpha\beta} =
      [\Delta^{\mu}_{\alpha}\Delta^{\nu}_{\beta} +
        \Delta^{\mu}_{\beta}\Delta^{\nu}_{\alpha} -
        (2/3)\Delta^{\mu\nu}\Delta_{\alpha\beta}]/2$, represents the
      traceless symmetric projection orthogonal to $u^{\mu}$.

The second law of thermodynamics, $\partial_{\mu}S^{\mu}\ge 0$, is
guaranteed to be satisfied if linear relationships between
thermodynamical fluxes and extended thermodynamic forces are imposed.
This leads to the following evolution equations for bulk and shear
\begin{align}
\Pi &= -\zeta\left[ \theta 
+ \beta_0 \dot \Pi 
+ \frac{4}{3}\beta_0 \theta\Pi \right], \label{bulk} \\ 
\pi^{\mu\nu} &= 2\eta\left[ \sigma^{\mu\nu} 
- \beta_2\dot\pi^{\langle\mu\nu\rangle} 
- \frac{4}{3}\beta_2 \theta\pi^{\mu\nu} \right] , \label{shear}
\end{align}
respectively,
where the coefficients of bulk and shear viscosity satisfy $\zeta,\eta
\ge 0$. The bulk and shear relaxation times, defined as $\tau_{\Pi} =
\zeta \beta_0$ and $\tau_{\pi} = 2 \eta \beta_2$, can be obtained
directly from the transport coefficients $\beta_0$ and $\beta_2$ which
are determined explicitly in the above derivations.

For Case 1, the coefficients $\beta_0$ and $\beta_2$ become
\begin{equation}\label{betas1}
\beta_0^{(1)} = 1/P,\quad  
\beta_2^{(1)} = 3/(\epsilon+P) + m^2\beta^2P/[2(\epsilon+P)^2],
\end{equation}
whereas for Case 2, they reduce to
\begin{equation}\label{betas2}
\beta_{0}^{(2)} = \frac{18}{5(\epsilon+P)} + \frac{3m^2\beta^2P}{5(\epsilon+P)^2},\quad
\beta_{2}^{(2)} = \beta_2^{(1)}.
\end{equation}
We note that although the relaxation time corresponding to shear
($\beta_2$) is the same for both the cases, that corresponding to bulk
($\beta_0$) is different. We stress that these coefficients have been
obtained consistently within a single theoretical framework. This is
in contrast to the standard derivation \cite {Israel:1979wp}, where
the transport coefficients have to be estimated from an alternate
theory.


\section{Thermal dilepton and hadron production}
 
Particle production is influenced by viscosity in two ways: first
through the viscous hydrodynamic evolution of the system and second
through corrections to the particle production rate via the
non-equilibrium distribution function
\cite{Teaney:2003kp}. Hydrodynamic evolution was considered in the
previous section; here we will concentrate on the thermal dilepton and
hadron production rates in heavy-ion collisions. While the hadrons are
emitted mostly in the final stages of the evolution, the dileptons are
emitted at all stages and thus probe the entire temperature history of
the system.

In the quark-gluon plasma (QGP) phase, the dominant production
mechanism for dileptons is $q\bar q\rightarrow \gamma^* \rightarrow
l^+ l^-$, whereas in the hadronic phase the main contribution arises
from $\pi^+ \pi^-\rightarrow \rho^0 \rightarrow l^+ l^-$. The dilepton
production rate for these processes is given by \cite{rvogt}
\begin{align}
 \frac{dN}{d^4x d^4p}=g^2 \int & \frac{d^3\textbf{p}_1}{(2\pi)^3} \frac{d^3\textbf{p}_2}{(2\pi)^3} f(E_1,T) f(E_2,T) \nonumber \\
& \times v_{rel} \sigma(M^2)\delta^4(p-p_1-p_2),
\label{dpr}
\end{align}
where $p_i=(E_i,\textbf{p}_i)$ are the four-momenta of the incoming
particles having equal masses $m_i$ and relative velocity
$v_{rel}=M(M^2-4m_i^2)^{1/2}/2E_1E_2$. Further, $M$ and $\sigma(M^2)$
are the dilepton invariant mass and production cross section,
respectively. Substituting for $f=f_0+\delta f$ and retaining only the
terms linear in $\delta f$, the dilepton production rate can be
expressed as a sum of contributions due to ideal, shear, and bulk:
\begin{eqnarray}
 \frac{dN}{d^4xd^4p}
= \frac{dN^{(0)}}{d^4xd^4p} + \frac{dN^{(\pi)}}{d^4xd^4p} + \frac{dN^{(\Pi)}}{d^4xd^4p}. 
\label{dilrate}
\end{eqnarray}
For the case $M\gg T\gg m_i$, the equilibrium distribution functions
can be approximated by the Maxwell-Boltzmann form $f(E,T)= \exp(-E/T)$
and $v_{rel}\simeq M^2/2E_1E_2$. In the QGP phase (for $q\bar q$
annihilation) we have $M^2g^2\sigma(M^2)=(80\pi/9)\alpha^2$ (with
$N_f$=2 and $N_c=3$) and in the hadronic phase (for $\pi^+\pi^-$
annihilation) we have $M^2g^2\sigma(M^2)=(4\pi/3)\alpha^2
|F_\pi(M^2)|^2$ \cite{rvogt}. The electromagnetic pion form factor is
$|F_\pi(M^2)|^2=m_\rho^4/[(m_\rho^2-M^2)^2+m_\rho^2 \Gamma_\rho^2]$,
where $m_\rho=775$ MeV and $\Gamma_\rho=149$ MeV are the mass and
decay width of the $\rho(770)$ meson, respectively \cite{Song:2010fk}.

With the above approximations, the integrals in Eq. (\ref{dilrate})
can be performed. The ideal part is given by \cite{rvogt}
\begin{equation}
  \frac{dN^{(0)}}{d^4xd^4p}=\frac{1}{2}~\frac{M^2g^2\sigma(M^2)}{(2\pi)^5}~e^{-p_0/T}.\label{dilrate0}
\end{equation}
The shear viscosity contribution is the same for $\phi_1$ and
$\phi_2$, Eqs. (\ref{phi1}) and (\ref{phi2}), and is given by
\cite{Dusling:2008xj}
\begin{equation}
\frac{dN^{(\pi)}}{d^4xd^4p}=
\frac{2}{3}\left(\frac{p^\mu p^\nu}{2sT^3}\pi_{\mu\nu} \right)\frac{dN^{(0)}}{d^4xd^4p},
\label{ratesh}
\end{equation}
where $s=(\epsilon+P)/T$ is the equilibrium entropy density. The bulk
viscosity contribution for $\phi_1$ is
\begin{equation}
\frac{dN^{(\Pi)}_1}{d^4xd^4p}
=\frac{\Pi}{P}\,\frac{dN^{(0)}}{d^4xd^4p},\label{rate1}
\end{equation}
and that for $\phi_2$ can be expressed as \cite{Bhatt:2011kx}
\begin{equation}
\frac{dN^{(\Pi)}_2}{d^4xd^4p}=
\frac{2}{5 sT^3} \left(\frac{M^2}{12}g^{\alpha\beta}-\frac{1}{3} p^\alpha p^\beta \right) \Delta_{\alpha\beta}\Pi \, \frac{dN^{(0)}}{d^4xd^4p}.
\label{rate2}
\end{equation}


The hadron spectra are obtained using the Cooper-Frye freezeout
prescription \cite{Cooper:1974mv}
\begin{equation}
\frac{dN}{d^2p_Tdy} = \frac{g}{(2\pi)^3} \int p_\mu d\Sigma^\mu f(x,p),
\label{CF}
\end{equation}
where $d\Sigma^\mu$ represents the element of the three-dimensional
freezeout hypersurface, and $f(x,p)$ represents the phase-space
distribution function at freezeout.

For the two cases discussed above we shall study the evolution of the
hydrodynamic variables and their influence on the dilepton and
hadron production rates.


\section{Bjorken scenario}

We consider the evolution of the system in longitudinal scaling
expansion at vanishing net baryon number density
\cite{Bjorken:1982qr}. In terms of the Milne coordinates
($\tau,r,\varphi,\eta_s$), where $\tau = \sqrt{t^2-z^2}$ and
$\eta_s=\tanh^{-1}(z/t)$, and with $u^\mu=(1,0,0,0)$, evolution
equations for $\epsilon$, $\Phi \equiv -\tau^2 \pi^{\eta_s \eta_s}$,
and $\Pi$ become
\begin{align}
\frac{d\epsilon}{d\tau} &= -\frac{1}{\tau}\left(\epsilon + P + \Pi -\Phi\right), \label{BED} \\
\tau_{\pi}\frac{d\Phi}{d\tau} &= \frac{4\eta}{3\tau} - \Phi - \frac{4\tau_{\pi}}{3\tau}\Phi, \label{Bshear} \\
\tau_{\Pi}\frac{d\Pi}{d\tau} &= -\frac{\zeta}{\tau} - \Pi - \frac{4\tau_{\Pi}}{3\tau}\Pi \label{Bbulk}.
\end{align}
The bulk and shear relaxation times, $\tau_{\Pi} = \zeta \beta_0$ and
$\tau_{\pi} = 2 \eta \beta_2$, reduce to
\begin{equation}\label{Mtaus}
\tau_{\Pi}^{(1)} = \frac{\epsilon+P}{PT} \left(\frac{\zeta}{s} \right), \quad
\tau_{\Pi}^{(2)} = \frac{18}{5T}\left(\frac{\zeta}{s}\right), \quad
\tau_{\pi} = \frac{6}{T}\left(\frac{\eta}{s}\right),
\end{equation}
for the two different forms of $\phi$, Eqs. (\ref{phi1}) and
(\ref{phi2}).

Once the temperature evolution is known from the hydrodynamical model,
the total dilepton spectrum is obtained by integrating the total rate
over the space-time evolution of the system
\begin{align}\label{total}
\frac{dN_{1,2}}{d^2p_TdM^2dy}= \pi R_A^2
\int_{\tau_0}^{\tau_{fo}}
d\tau ~\tau \int_{-\infty}^{\infty}d\eta_s
\left(\frac{1}{2}\frac{dN_{1,2}}{d^4xd^4p}\right),
\end{align}
where $\tau_0$ and $\tau_{fo}$ are the initial and freezeout times for
the hydrodynamic evolution. We note that for the Bjorken expansion,
$d^4x=\pi R_A^2d\eta_s\tau d\tau$, where $R_A=1.2 A^{1/3}$ is the
nuclear radius.

In $(\tau,r,\varphi,\eta_s)$ coordinates, particle four-momentum is
$p^{\mu} = (m_T \cosh(y-\eta_s),~p_T \cos(\varphi_p- \varphi),~p_T
\sin (\varphi_p-\varphi)/r,~m_T \sinh(y-\eta_s)/\tau)$, where $m_T^2$
= $p_T^2+m^2$. The other factors appearing in the rate expressions,
Eqs. (\ref{ratesh}) and (\ref{rate2}), are then given by
\begin{eqnarray}\label{visc-fcts1}
 p^{\alpha}p^{\beta} \pi_{\alpha\beta} 
&=& \frac{\Phi}{2} p_T^2-\Phi ~m_T^2~\sinh^2(y-\eta_s),\\
 p^\alpha p^\beta \Delta_{\alpha\beta} 
&=& -p_T^2-m_T^2~\sinh^2(y-\eta_s). \label{visc-fcts2}
\end{eqnarray}

Similar to the dilepton spectra, the hadronic spectra can also be
split up into three parts. Writing the momentum flux through the
hypersurface element as $p_\mu d\Sigma^\mu = m_T \cosh(y-\eta_s) \tau
d\eta_s r dr d\varphi$, and performing the $\eta_s$ integration, we
get for the ideal case,
\begin{equation}
\frac{dN^{(0)}}{d^2p_Tdy} = \frac{g}{(2\pi)^2}m_T\tau_{fo} R_A^2 K_1(z_m),
\end{equation}
where $K_n$ are the modified Bessel functions of the second kind and
$z_m\equiv m_T/T$. The contribution due to the shear viscosity to the
hadron production reduces to
\begin{equation}
\frac{dN^{(\pi)}}{d^2p_Tdy} = \frac{\Phi}{4(\epsilon+P)} \left[z_p^2 - 2z_m \frac{K_2(z_m)}{K_1(z_m)}\right]\frac{dN^{(0)}}{d^2p_Tdy},
\end{equation}
where $z_p\equiv p_T/T$. The bulk viscosity contribution in Case 1,
Eq. (\ref{phi1}), is calculated to be
\begin{equation}
\label{hrate1}
\frac{dN^{(\Pi)}_1}{d^2p_Tdy} = \frac{\Pi}{P}\frac{dN^{(0)}}{d^2p_Tdy} ,
\end{equation}
whereas in Case 2, Eq. (\ref{phi2}), it reduces to
\begin{equation}
\label{hrate2}
\frac{dN^{(\Pi)}_2}{d^2p_Tdy} =  \frac{-\Pi}{5(\epsilon+P)}\left[z_p^2 + z_m \frac{K_2(z_m)}{K_1(z_m)}\right]\frac{dN^{(0)}}{d^2p_Tdy}.
\end{equation}
Here we have used the recurrence relation
$K_{n+1}(z)=2nK_n(z)/z+K_{n-1}(z)$. It is important to note that the
bulk viscosity contribution in Case 1 is negative, whereas that in
Case 2 is positive ($\Pi<0$).

\section{Numerical results and discussion}

We now present numerical results for the Bjorken expansion of the
medium for the initial temperature $T_0=310$ MeV and time $\tau_0=0.5$
fm/c, typical for the Relativistic Heavy-Ion Collider.  The freezeout
temperature was taken to be $160$ MeV. The initial pressure configuration
was assumed to be isotropic: $\Phi=0=\Pi$. We employ the equation of
state of Refs. \cite{Huovinen:2009yb,Bazavov:2009zn} based on a recent
lattice QCD simulation. The shear viscosity to entropy density ratio
$\eta/s$ was taken to be $1/4\pi$ corresponding to the conjectured
lower bound obtained in Ref. \cite{Kovtun:2004de}. For the bulk
viscosity to entropy density ratio $\zeta/s$ at $T \geq T_c \approx
184$ MeV, we adopted a parametrized form of the lattice QCD result; see
Refs. \cite{Meyer:2007dy,Rajagopal:2009yw}.  For $T<T_c$, we
parametrized $\zeta/s$ given in Ref. \cite{Prakash:1993bt}.

\begin{figure}[t]\begin{center}
\scalebox{0.33}{\includegraphics{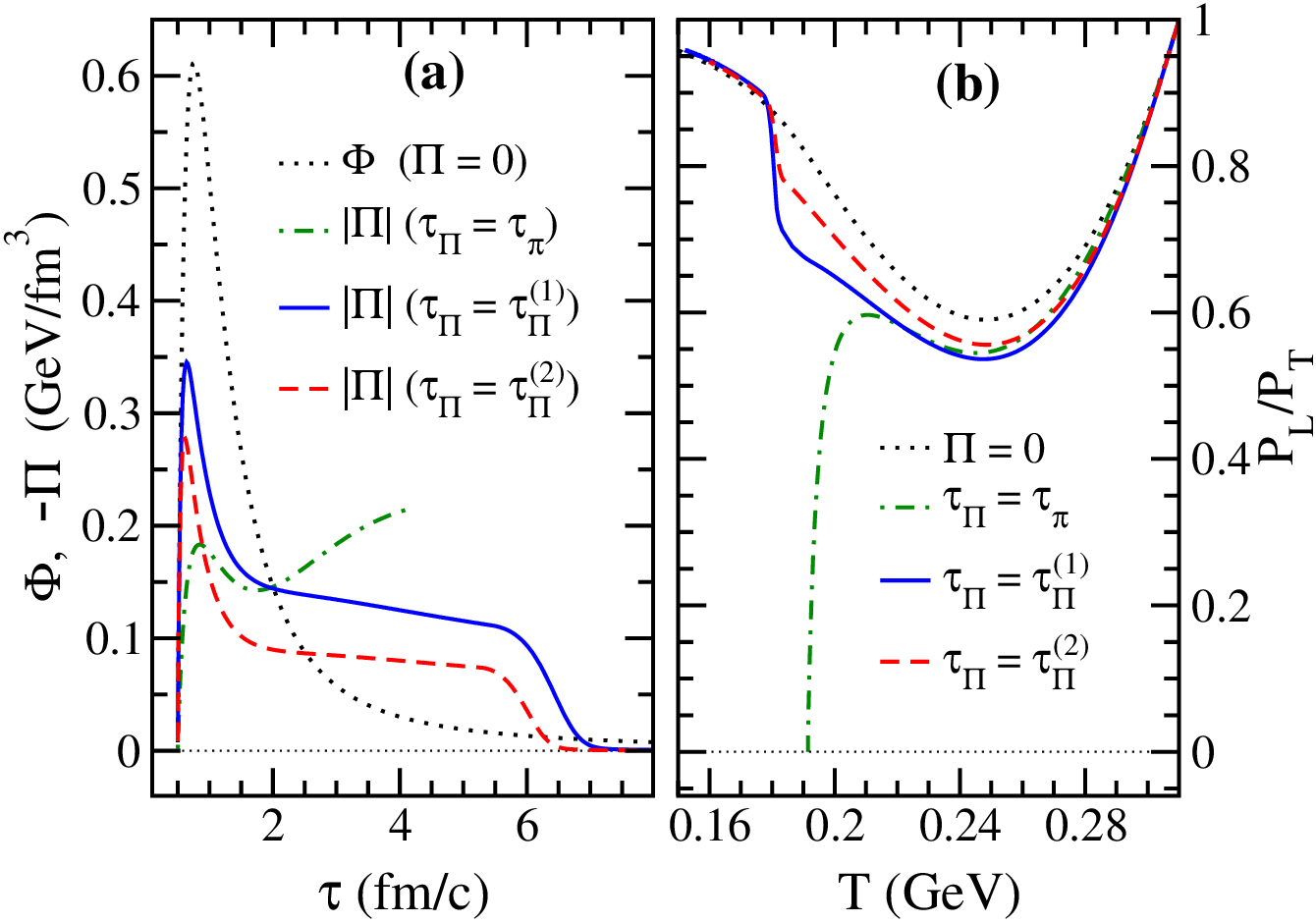}}
\end{center}
\vspace{-0.4cm}
\caption{(color online) (a) Time evolution of shear $\Phi$ and bulk
  $\Pi$ viscous pressures, and (b) temperature dependence of the ratio
  of the longitudinal to transverse pressures $P_L/P_T$, for the
  various bulk relaxation times $\tau_\Pi$ defined in
  Eq. (\ref{Mtaus}). Note that for $\tau_\Pi=\tau_\pi$, cavitation
  ($P_L<0$) sets in.}
\label{plpt}\end{figure}

Figure \ref{plpt}(a) presents the time evolution of shear $\Phi$ and
bulk $\Pi$ viscous pressures for the various bulk relaxation times
$\tau_\Pi$ defined in Eq. (\ref{Mtaus}). At times $\tau \gtrsim 3$
fm/c, corresponding to temperatures $T \lesssim 1.2 ~T_c$, the bulk
dominates the shear pressure which can influence the particle
production appreciably. The widely used choice $\tau_\Pi = \tau_\pi$
(dotted-dashed line) leads to vanishing longitudinal pressure $P_L$ and
cavitation \cite{Rajagopal:2009yw} as is evident in
Fig. \ref{plpt}(b). On the other hand, $\tau_\Pi=\tau_\Pi^{(1,2)}$
does not lead to cavitation as discussed in \cite{Jaiswal:2013fc}. As
$\tau_\Pi^{(1)} > \tau_\Pi^{(2)}$ at all times, the magnitude of $\Pi$
is found to be larger in Case 1 (solid line). This leads to enhanced
pressure anisotropy, i.e., a larger departure of
$P_L/P_T=(P+\Pi-\Phi)/(P+\Pi+\Phi/2)$ from unity.

\begin{figure}[t]\begin{center}
\scalebox{0.3}{\includegraphics{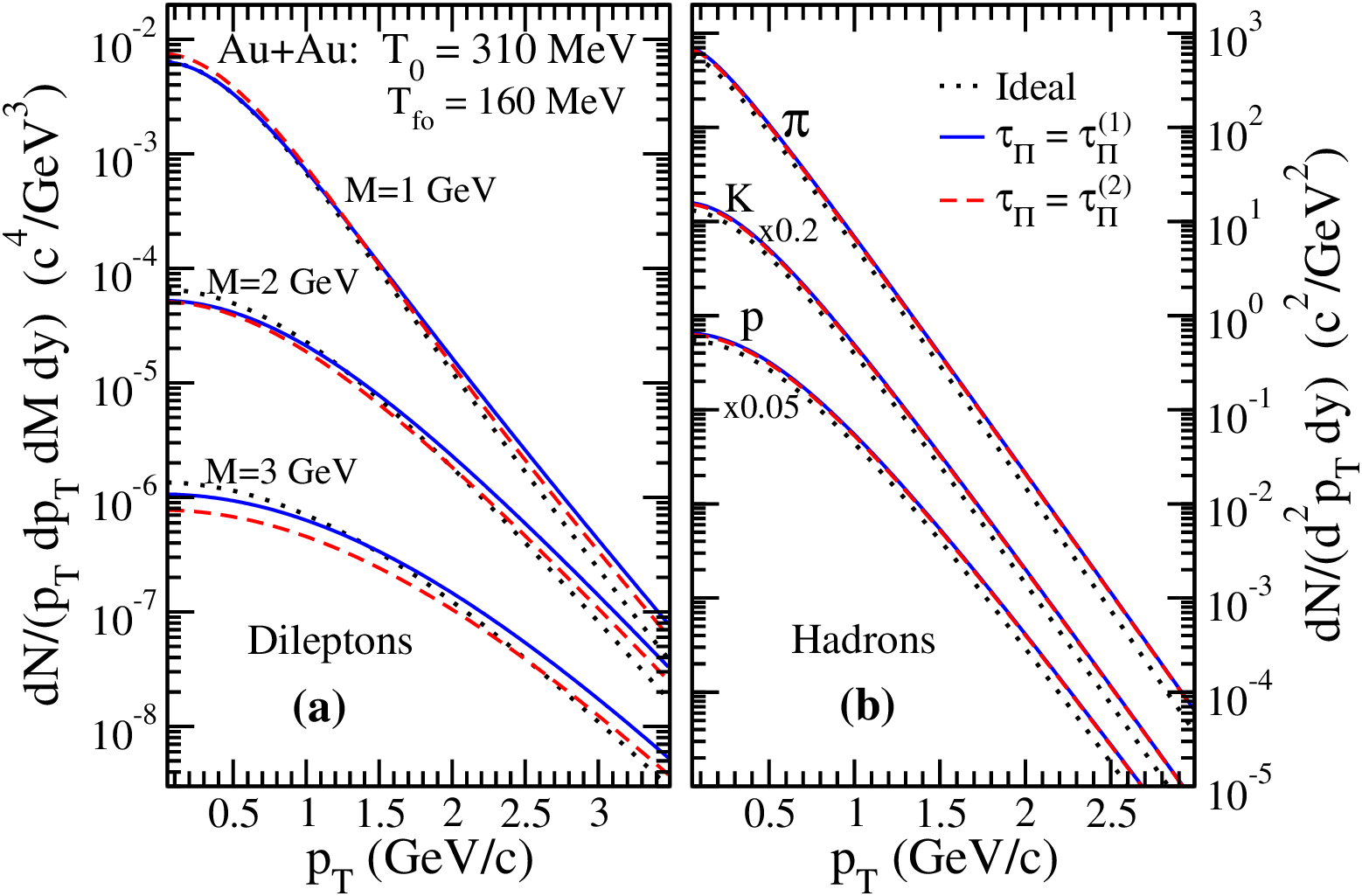}}
\end{center}
\vspace{-0.4cm}
\caption{(color online) Particle spectra as a function of the
  transverse momentum $p_T$, for ideal and viscous hydrodynamics with
  bulk relaxation times $\tau_\Pi$ defined in Eq. (\ref{Mtaus}) for
  (a) dileptons of invariant mass $M=1,~2,~3$ GeV/$c^2$, and (b)
  hadrons.}
\label{dnpt}\end{figure}

\begin{figure}[b]\begin{center}
\scalebox{0.35}{\includegraphics{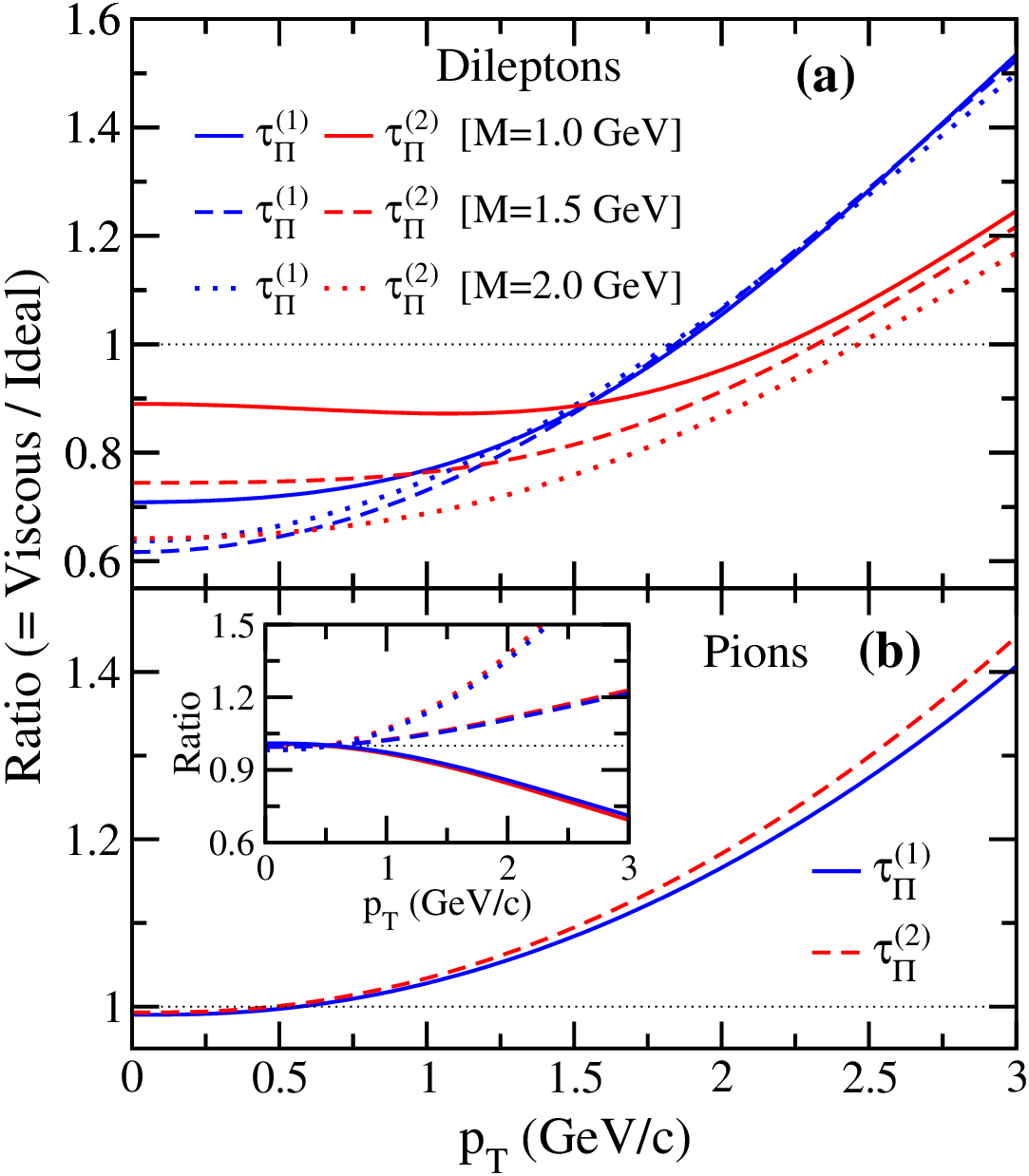}}
\end{center}
\vspace{-0.4cm}
\caption{(color online) Ratios of particle yields for viscous and
  ideal hydrodynamics as a function of $p_T$, for the two bulk
  relaxation times $\tau_\Pi$ defined in Eq. (\ref{Mtaus}) for (a)
  dileptons of invariant mass $M=1,~1.5,~2$ GeV/$c^2$, and (b)
  pions. Inset: Pion yields in various evolution and production
  scenarios scaled by the consistent second-order calculation for Case
  1 [blue (dark grey)] and Case 2 [red (light grey)]. 
  Solid lines: second-order evolution with
  ideal production rate; Dashed lines: second-order evolution with
  first-order correction to the production rate; Dotted lines: ideal
  evolution with first-order correction to the production rate.}
\label{ratio}\end{figure}

Figure \ref{dnpt} displays dilepton and hadron transverse momentum
spectra for the two choices of $\tau_\Pi$, in comparison with the
ideal hydrodynamic calculation, and Fig. \ref{ratio} shows the same
spectra normalized by the ideal case. Note the enhancement of the
dilepton spectra at high $p_T$, and their suppression at low $p_T$ compared
to the ideal case. The high-$p_T$ dileptons emerge predominantly at
early times when the temperature and density are high. Viscosity slows
down the cooling of the system \cite{Muronga:2003ta} producing
a relatively larger number of hard dileptons. We observe that at high
$p_T$, the viscous correction to the dilepton production rate due to
shear is positive and dominates that due to bulk. The low-$p_T$
dileptons are produced mainly at later stages of the evolution when
the negative correction due to the bulk viscosity dominates
(Fig. \ref{plpt}) leading to the suppression of the spectra compared
to the ideal case. Further for Case 2 [red (light grey lines)], the $p_T^2$
dependence of the viscous correction, Eqs. (\ref{rate2}) and
(\ref{visc-fcts2}), implies a smaller enhancement (suppression) at
high (low) $p_T$, compared to Case 1 [blue (dark grey lines)]. 
The $M$-dependent
splitting is consistent with Eqs. (\ref{rate1}) to (\ref{rate2}).

Figure \ref{ratio}(b) shows the pion spectra scaled by the ideal case
for the two choices of $\tau_\Pi$. The negative contribution from the
bulk viscous correction for Case 1, Eq. (\ref{hrate1}), causes
suppression of the ratio relative to Case 2, Eq. (\ref{hrate2}), where
the correction is positive. More massive hadrons display qualitatively
similar behavior. Interestingly, at high $p_T$, dileptons and hadrons
display opposite trends for $\tau_\Pi^{(1)}$ and $\tau_\Pi^{(2)}$
(Fig. \ref{ratio}).

\begin{figure}[t]\begin{center}
\scalebox{0.34}{\includegraphics{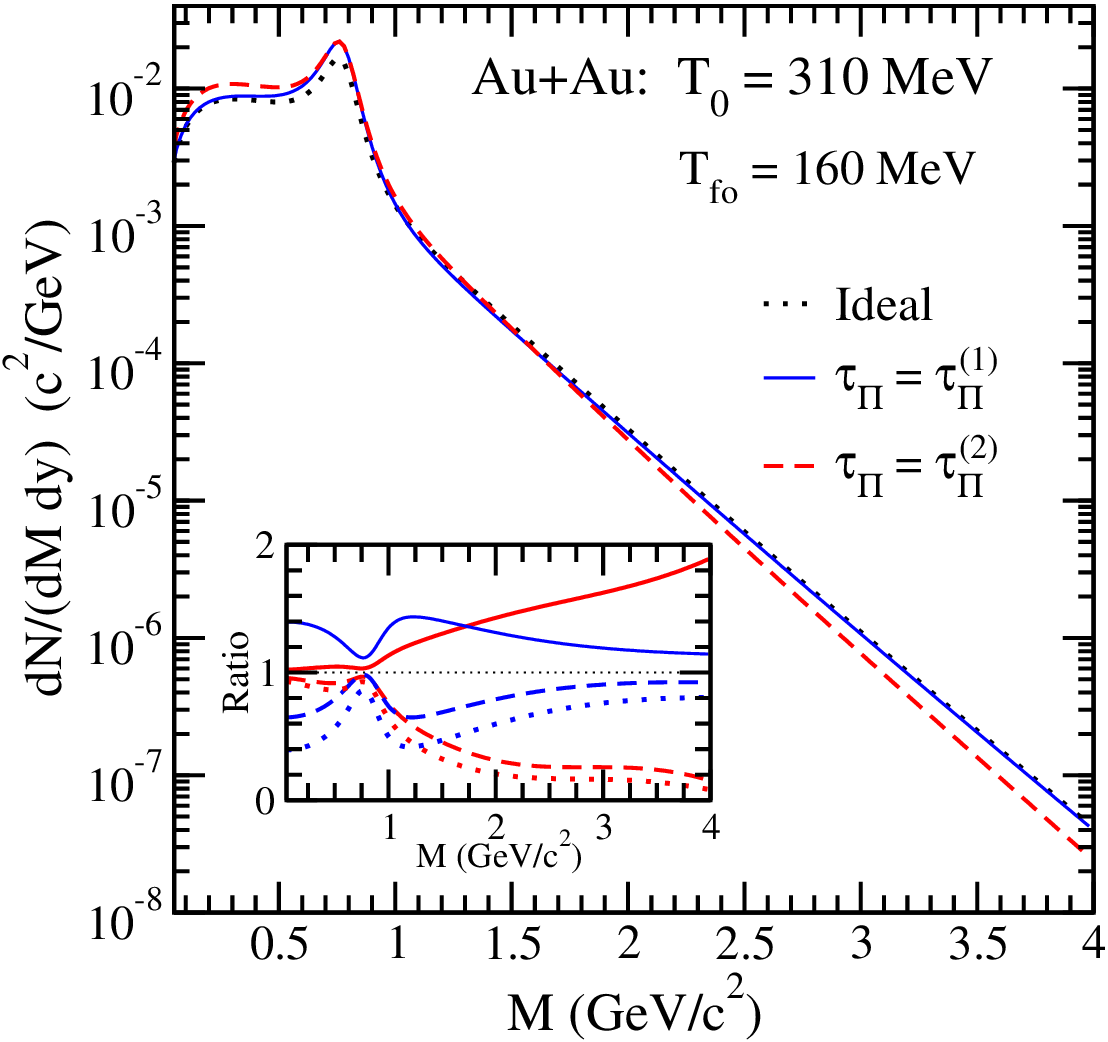}}
\end{center}
\vspace{-0.4cm}
\caption{(color online) Dilepton yields as a function of the invariant
  mass $M$, in ideal and viscous hydrodynamics with the two bulk
  relaxation times $\tau_\Pi$ defined in Eq. (\ref{Mtaus}). Inset:
  Same as Fig. \ref{ratio} inset but for dileptons.}
\label{dndm}\end{figure}

Finally, Fig. \ref{dndm} shows the dilepton invariant mass spectra for
the two cases [Eqs. (\ref{phi1}) and (\ref{phi2})] in comparison with the
ideal case. Results based on Case 1 (blue solid) are almost the same
as those obtained in the ideal case at all invariant masses. This is
essentially due to the fact that the invariant mass spectrum is
dictated by the yields at small $p_T$ where the two are nearly
identical (Fig. \ref{dnpt}(a)). For Case 2 (red dashed) the spectrum
exhibits enhanced low-mass and suppressed high-mass dilepton
yields. This again can be traced back to the trend seen in
Fig. \ref{dnpt}(a). Note that the peak at $M=0.77$ GeV/$c^2$ corresponds to
the dilepton production from the $\rho(770)$ decay.

In contrast to the consistent approach adopted here, in
Refs. \cite{Monnai:2009ad,Dusling:2008xj}, ideal hydrodynamical
evolution was followed by particle production with non-ideal $f(x,p)$
up to first order in gradients. On the other hand, in
Refs. \cite{Bhatt:2011kx,Denicol:2009am,Dusling:2011fd}, although the
evolution was according to the second-order viscous hydrodynamics, the
freezeout procedure involved ideal \cite{Denicol:2009am} or
Navier-Stokes \cite{Bhatt:2011kx,Dusling:2011fd} corrections to the
$f(x,p)$. To illustrate the differences arising due to inconsistent
approaches, we show, in the insets of Figs. \ref{ratio} and \ref{dndm},
pion and dilepton production rates, respectively,
in various evolution and production
scenarios scaled by the rate obtained in a consistent second-order
calculation. We find that the results deviate from unity significantly
which may have important implications for the on-going efforts to
extract transport properties of QGP within a hydrodynamic framework.

\section{Summary}

Starting with the entropy four-current expressed in terms of the
single-particle distribution function, and using two
different forms of the correction to the equilibrium distribution
function, we derived second-order evolution equations for the viscous
dissipative fluxes by invoking the generalized form of the second law
of thermodynamics. For consistency, the same non-equilibrium
distribution functions were used in the two calculations of particle
production. One of the two forms
(Eq. (\ref{phi1})) has been used here for the first time to study
particle production. In the Bjorken scaling expansion, appreciable
differences were found in the two sets of results. 

More importantly,
we also compared results of various inconsistent calculations with
those of a consistent second-order calculation; see insets in Figs.
\ref{ratio} and \ref{dndm}. It is clear that the particle yields are
significantly affected if the viscous effects in hydrodynamic
evolution and particle production are not mutually consistent. There
is no {\it a priori} reason to believe that in  realistic 2+1 or 3+1
dimensional
calculations this will not be the case. It will then have important
consequences for the extraction of transport properties of the
quark-gluon plasma in the hydrodynamic scenario. In conclusion, the
non-equilibrium distribution function used in the particle production
should be the same as that used in the derivation of hydrodynamic
evolution equations.

\end{document}